\documentclass[iop]{emulateapj}
\usepackage{amssymb}
\usepackage{amsmath}
\usepackage{float}

\def\be{\begin{equation}}
\def\ee{\end{equation}}
\def\ba{\begin{eqnarray}}
\def\ea{\end{eqnarray}}

\newcommand{\DM}{\ensuremath{\mathrm{DM}}} 
\newcommand{\DMhat}{\widehat{\rm DM}}
\newcommand{\Dtiss}{\Delta t_{\rm ISS}}
\newcommand{\DtissGHz}{\Delta t_{\rm ISS,GHz}}
\newcommand{\q}{\rm q}

\newcommand{\SFphi}{D_\phi}
\newcommand{\SFDM}{D_{\DM}}
\newcommand{\xvec}{{\bf x}}
\newcommand{\bvec}{{\bf b}}

\newcommand{\veffperp}{{v_{\rm eff_{\perp}}}}

\shorttitle{Dispersion Measure Errors in Pulsar Timing}
\shortauthors{Lam et al.}

\begin{document}

\title{Pulsar Timing Errors from Asynchronous Multi-frequency Sampling of Dispersion Measure Variations}

\author{M. T. Lam, J. M. Cordes, S. Chatterjee, T. Dolch}
\affil{Department of Astronomy, Cornell University, Ithaca, NY 14853; mlam@astro.cornell.edu}

\begin{abstract}
Free electrons in the interstellar medium cause frequency-dependent delays in pulse arrival times due to both scattering and dispersion. Multi-frequency measurements are used to estimate and remove dispersion delays. In this paper, we focus on the effect of any non-simultaneity of multi-frequency observations on dispersive delay estimation and removal. Interstellar density variations combined with changes in the line-of-sight from pulsar and observer motions cause dispersion measure variations with an approximately power-law power spectrum, augmented in some cases by linear trends.   We simulate time series, estimate the magnitude and statistical properties of timing errors that result from non-simultaneous observations, and derive prescriptions for data acquisition that are needed in order to achieve a specified timing precision. For nearby, highly stable pulsars,  measurements need to be simultaneous to within about one day in order that the timing error from asynchronous DM correction is less than about 10~ns. We discuss how timing precision improves when increasing the number of dual-frequency observations used in dispersion measure estimation for a given epoch. For a Kolmogorov wavenumber spectrum, we find about a factor of two improvement in precision timing when increasing from two to three observations but diminishing returns thereafter.

\end{abstract}


\section{Introduction}

One of the goals of precision pulsar timing is the detection of low-frequency ($\sim$~nanohertz) gravitational waves (GWs) from sources at cosmological distances and possibly from Galactic sources \citep{Detweiler1979,hd1983,Chernoff2009,Sesana2013}. Such a detection requires sub-microsecond timing precision, which is challenging due to a variety of astrophysical and  instrumentation effects that must either be  mitigated or fitted for in timing models \citep[e.g.][]{Jenet+2005}.   The measurement model for pulse times-of-arrival (TOAs) includes: (i) deterministic contributions from spin kinematics, orbital motions, and interstellar propagation delays; (ii) stochastic timing noise from pulsars themselves and from the interstellar medium; and (iii) measurement noise  \citep{rt1991,sac+1998,lk2012}.

The ionized interstellar medium (ISM) induces various frequency-dependent effects on TOAs from dispersion, refraction, and scattering \citep{Cordes2013,Stinebring2013}.   Any such effects that are larger than the  TOA precision required for GW detection need to be removed by using multiple-frequency observations. In this paper, we consider some of the requirements for removing the  dispersive delay, which is the largest frequency-dependent interstellar effect. For a cold, unmagnetized plasma, a pulse observed at radio frequency $\nu$ is delayed compared to one at infinite frequency by an amount $t_{\DM} \propto \DM/\nu^{2}$, where the dispersion measure (\DM) is the line-of-sight (LOS) integral of the electron density. \DM\ is epoch-dependent because the LOS changes from motions of the pulsar and the Earth and because of turbulent and bulk motions within  the ISM itself \citep{pw1991,cr1998}. Therefore, the dispersion delay  must be removed on an epoch-by-epoch basis.  Measurements at two or more frequencies are used to estimate \DM\ and then subtract the dispersion delay to obtain infinite-frequency TOAs that are intended to be devoid of interstellar plasma delays.

Pulsar timing arrays (PTAs) are ensembles of recycled, millisecond pulsars (MSPs) that can potentially provide high precision TOAs usable for GW detection. Currently, TOAs are typically obtained in observing campaigns with  a roughly monthly cadence between epochs \citep{dfg+2013,Hobbs2013,kc2013}.  However, TOAs may be measured at individual frequencies over a period of several days around each of these epochs.  Variations in DM over this time range can contaminate the estimated DM and consequently also the infinite-frequency TOAs.   Though variations in DM are small over periods of days ($\Delta\DM / \DM \sim 10^{-4} - 10^{-5}$), they are large enough to add significantly to the timing error budget. 

The North American Nanohertz Observatory for Gravitational Waves (NANOGrav; \citealt{McLaughlin2013}) observes MSPs using two facilities, the Arecibo Observatory and the Green Bank Telescope (GBT). Most pulsars in the PTA are observed roughly once a month for 20-30 minutes  in each of two frequency bands, chosen on a per-pulsar basis to optimize the precision of the DM correction. The mechanical agility of the receiver turret at Arecibo allows two frequency bands to be observed sequentially on the same day. The system at the GBT requires physical switching between receivers at different foci of the telescope used for each frequency. This switching must be done on a pulsar-by-pulsar basis and is time-inefficient. Instead, pulsars are observed at both frequencies on separate days, resulting in gaps between observations ranging from roughly one day to a week. These gaps have been mitigated in the NANOGrav processing pipeline by combining observations in 15-day wide bins and evaluating DM with a piecewise constant model \citep{dfg+2013}  that is fitted to the multifrequency data within each bin under the assumption that DM is constant over 15 days.   
 
There are additional frequency-dependent effects that we do not address in this work, including refraction and scattering in the ISM  \citep{cpl1986,rnb1986,Rickett1990,fc1990} and  frequency-dependent variations of pulse shapes that are intrinsic to each pulsar \citep{Craft1970,hr1986,kll+1999,hsh+2012,Pennucci+2014}. Other interstellar effects on pulsar timing are also being investigated. J. M. Cordes et al. (in preparation) analyze the dependence of DM on frequency that results from spatial averaging due to multipath scattering.   In another (M. T. Lam et al., in preparation), we diagnose contributions to DM variations that result from both the change in pulsar distance and from the change in direction of the LOS. Here we focus on the timing uncertainties that result specifically from the non-simultaneity of the multi-frequency observations. Our work gives an exact treatment over the analysis presented in \cite{cs2010}.

In \S 2, we present the mathematical framework of the effect of non-simultaneous observations on DM estimation and the associated timing errors. In \S 3, we describe our simulations of DM time series and in \S 4 we present the results of these simulations. We conclude in \S 5 by describing the overall impact on the timing noise budget and GW sensitivity.

\section{Timing Errors from DM Mis-estimation}

\newcommand{\dTOA}{\Delta t}

Consider TOA measurements made at two epochs $t_1 = t$ and $t_2 = t + \tau$  at frequencies $\nu_1$ and $\nu_2$, respectively. For specificity, we assume $\nu_1 > \nu_2$. A perfect timing model would allow the pulse arrival time to be predicted with zero error. However, timing perturbations are expected from both achromatic effects (e.g. due to GWs and from errors in the pulsar spin or orbital parameters) and chromatic, interstellar effects.  Chromatic cold plasma effects always decrease with increasing frequency,  so TOAs are referenced to infinite frequency. Defining $\dTOA_{\infty}$ as the achromatic, infinite frequency perturbation, we write the total timing perturbation from both achromatic effects and dispersion as
\be 
\dTOA_{i} = \dTOA_{\infty}(t_{i}) +  K \nu_{i}^{-2} \DM(t_{i}),
\ee
where the subscript $i=1,2$ denotes the epoch and $K \equiv c r_e / 2 \pi \approx 4.149  \mathrm{\;ms\; GHz^2\;pc^{-1}\;cm^3}$ is the dispersion constant in observationally convenient units,  with $c$ the speed of light and $r_e$ the classical electron radius \citep{lk2012}. For simultaneous observations ($\tau = 0$), $\DM(t_{1,2})$ is  constant  and $\dTOA_\infty(t)$ can be solved for exactly assuming there is no  measurement error.

For non-simultaneous observations ($\tau \ne 0$), estimation of DM and correction to infinite frequency will be in error according to the change in actual DM between the two epochs.   If the difference in TOA offsets is attributed {\em solely} to dispersion delays with a fixed value of \DM\ and if the achromatic offset $\dTOA_\infty$ is the same at the two epochs, the estimated DM fluctuation\footnote{To simplify notation, we will assume that the average DM has been removed from all DM time series.} at epoch $t$ is
\ba
\!\!\!\!\!\!
\DMhat(t, \tau) &=& \frac{\dTOA_1 - \dTOA_2}{K \left(\nu_1^{-2} - \nu_2^{-2}\right)}
	= \frac{\DM(t) - r^2 \DM(t+\tau)}{1 - r^2}.
\label{eq:DMhat}
\ea
for a frequency ratio  $r = \nu_1 / \nu_2$. We use  the \DM\ increment over the interval between observations, 
\ba
 \Delta \DM (t, \tau) \equiv \DM(t) - \DM(t+\tau),
 \label{eq:dDM}
 \ea
 to express the difference between  true and estimated \DM\ as 
\ba
\delta \DMhat(t,\tau) & \equiv & \DM(t) - \DMhat(t,\tau) 
 = \frac{r^2\Delta\DM(t, \tau)}{r^2-1}.
\label{eq:deltaDM}
\ea

When the (mis)estimated  DM in Equation \eqref{eq:DMhat} is used to correct the measured TOA at epoch $t$ to infinite frequency, the systematic error is 
\be
\delta \hat{t}_\infty(t,\tau) = K \nu_1^{-2}\, \delta \DMhat(t,\tau)
\label{eq:toa_error}
\ee
The TOA error vanishes for $\tau = 0$. However, it has the curious property of a decline with increasing $r$ but is asymptotic to a constant value as $r$ goes to infinity. Note that flipping $\nu_1$ and $\nu_2$ will cause a change in $\delta\DMhat$ but not $\delta\hat{t}_\infty$. 

\section{DM Variations from ISM Structure}

Epoch-dependent DM variations are well known, e.g.\ for the Crab Pulsar \citep{ir1977}, for B1937+21 \citep{rtd1988,cwd+1990,ktr1994,rpb+2006}, for B1821--24 \citep{cl1997}, and numerous other cases \citep{pw1991,b+93,You+07,Keith+2013}. In some cases, $\DM(t)$ is consistent with sampling of stochastic electron-density variations while in others, linear trends in time are prominent.   

\begin{figure*}[t!]
\epsscale{1.2}
\begin{center}
\plotone{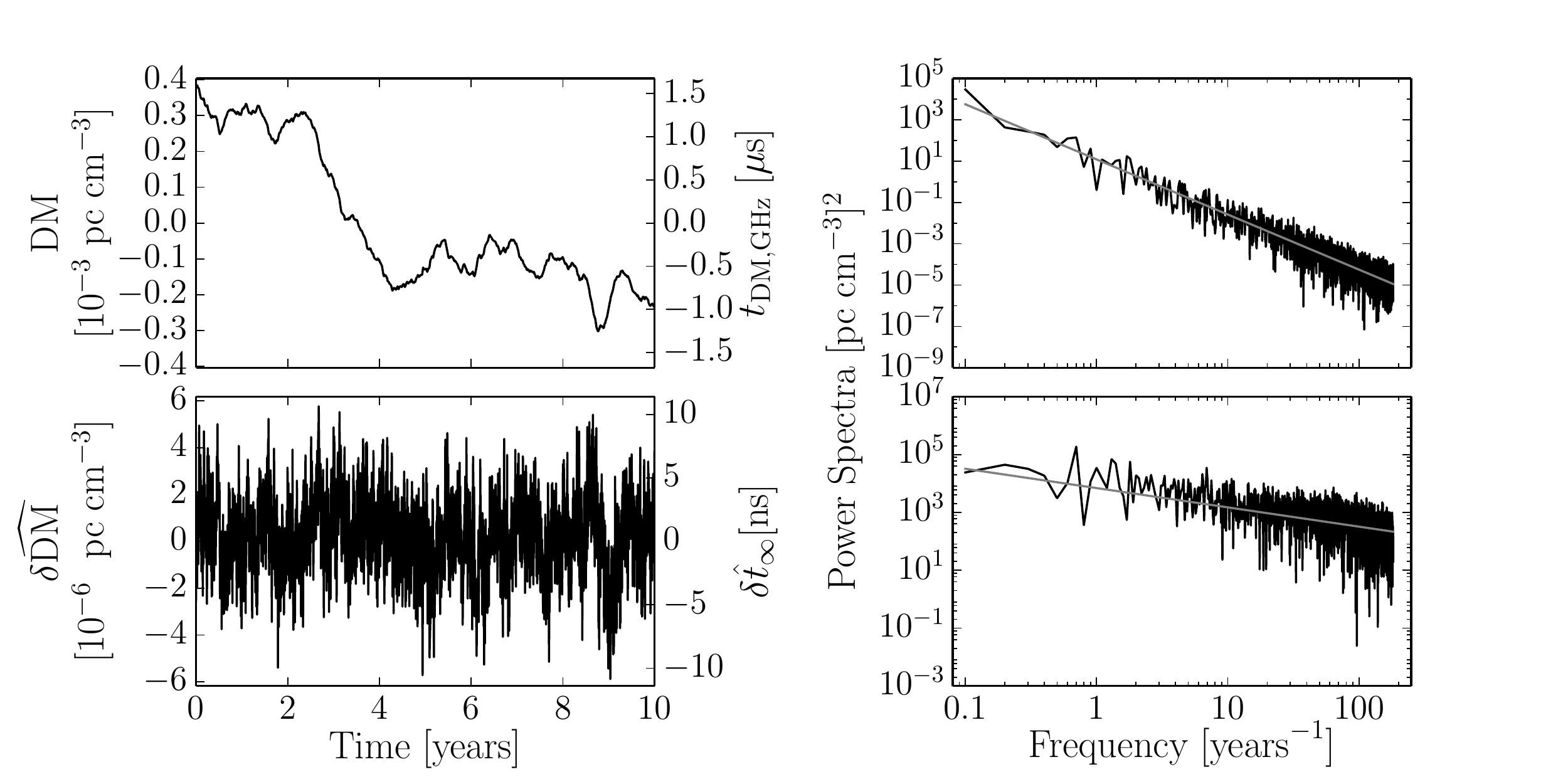}
\caption{\footnotesize 
A single realization of a simulated DM time series that results from density variations along the LOS in a medium with a Kolmogorov wavenumber spectrum and from changes in the LOS due to relative motions. The power spectrum for $\DM(t)$ scales as $f^{-\gamma}$ with $\gamma = 8/3$. 
Top left: $\DM(t)$ as a perturbation added to a mean DM with the left-hand scale in DM units and the right-hand scale giving the time delay $t_{\DM}$ in microseconds for $\nu = 1$~GHz.
Top right: Power spectrum of $\delta\DM(t)$ overplotted with a fitted straight line of slope $\gamma = 8/3$.  Bottom left: The difference between estimated and true DM, $\delta \widetilde{\DM}(t,\tau)$, for a lag $\tau =1~\mathrm{day}$ between dual-frequency measurements at $\nu_1=1.5$ and $\nu_2=0.8$ GHz with the right-hand scale set for the infinite-frequency TOA perturbation given by Equation \eqref{eq:toa_error}.
Bottom right: Power spectrum of the DM difference overplotted with a fitted straight line of slope $\gamma=2/3$.
\label{fig:samplerealization}
}
\end{center}
\end{figure*}

A linear trend in DM, modeled as $\DM(t) = \DM_0 + (d\DM/dt)t$, would give a timing error at frequency 
$\nu_1$ (in GHz)
\ba
\delta \hat{t}_\infty(t,\tau) &=& K \nu_1^{-2}\, (d\DM/dt)\tau
\nonumber \\
 &\approx& 1.14~{\rm ns}\, \nu_1^{-2} \tau_d \left( \frac{d\DM/dt}{10^{-4}~\rm {pc~cm^{-3}~yr^{-1}}} \right),
\label{eq:toa_error2}
\ea
where the approximate value is scaled to a nominal value of $d\DM/dt$ and with $\tau$ in days. Measured DM derivatives range from $d\DM/dt \approx 10^{-5}$~to $10^{-2}$~pc~cm$^{-3}$~yr$^{-1}$, so timing errors as large as $114$~ns will occur for lags of one day for LOSs with the largest DM derivatives.   However, a large linear trend is easy to recognize in timing data and a wide range of epochs can be used to estimate and remove it.    In the remainder of our analysis, we will therefore ignore the contribution from linear trends in DM and focus on stochastic variations. 

Fluctuations in $\DM(t)$ arise from density variations in the ISM that combine with the change in LOS from transverse  motions of the observer, pulsar, and medium\footnote{Linear trends result, in part, from radial motions along the LOS, so transverse motions are relevant to our discussion.}.  To describe electron-density variations, we use a power law wavenumber spectrum with cutoffs $q_1$ and $q_2$ and spectral coefficient $\mathrm{C_n^2}$,
\be
P_{\delta n_e}(\boldsymbol{q}) = \mathrm{C_n^2} q^{-\beta},~~~ q_1 \le q \le q_2,
\label{eq:wavenumber_spectrum}
\ee
that depends only on the magnitude of the wavenumber $q$, 
which applies to isotropic density irregularities consistent with many LOSs (see \citealt{bmg+2010} for evidence of anisotropic scattering towards B0834+06) . For $\beta>3$ and $\q_1 \ll \q_2$, the RMS electron density is dominated by the largest scales, $2\pi/q_1$. Kolmogorov turbulence is a benchmark model commonly used to describe fluctuations in the ISM consistent over many length scales (see \citealt{Rickett1990} for an overview). The Kolmogorov case corresponds to $\beta = 11/3$. 
Example time series are shown in Fig~\ref{fig:samplerealization}. 

Density fluctuations impose phase perturbations on electromagnetic waves that are manifested as variations in DM and as intensity variations (interstellar scintillations, ISS). The phase structure function (SF), $\SFphi(b) = \left\langle [\phi(\xvec) - \phi(\xvec+\bvec)]^2\right\rangle$,  is closely related to measurable ISS quantities \citep[e.g.][]{Rickett1990}. It scales as $\SFphi(b) \propto b^{\beta-2}$ for spatial separations $b$  intermediate between the smallest and largest scales in the ISM, $2\pi/q_2 \ll b \ll 2\pi/q_1$ along with  $2 < \beta  < 4$, which appears to be the range of $\beta$ that best characterizes ISS observations \citep{bcc+2004,lmg+2004}.  For brevity, we refer to this set of constraints as the `scintillation regime.' 

Measurements of the ISS timescale $\Dtiss$ in the strong scintillation regime correspond to  $\SFphi(\veffperp\Dtiss) = 1$~rad$^2$, 
where $\veffperp$ is a weighted combination of transverse velocities of the pulsar, observer, and ISM.  The phase structure function can be extrapolated to much longer time scales, subject to consistency with the above criteria on $b = \veffperp \tau$, using \citep[e.g.][]{fc1990},
\be
D_\phi(\tau) = (1~\mathrm{rad}^2)\left[\frac{\tau}{\Dtiss(\nu)}\right]^{\beta-2}.
\ee

DM variations are related to phase variations by $\delta\DM = -\nu \phi / c r_e$.  The time series $\delta\DM(t)$ is a red noise process with a power-law spectrum that scales as $S_{\DM}(f) \propto f^{-\gamma}$ where $\gamma = \beta - 1$ in the scintillation regime. The corresponding structure function for \DM, $D_{\DM}(\tau) = \left\langle\left[\Delta\DM(t,\tau)\right]^2 \right\rangle$, with $\Delta\DM(t,\tau)$ defined in Equation \eqref{eq:dDM},  is
\be
D_{\DM}(\tau) = \frac{D_\phi(\veffperp\tau)}{\left(\lambda r_e\right)^2} = \frac{\nu^2}{\left(c r_e \right)^2}\left[\frac{\tau}{\Dtiss(\nu)}\right]^{\beta-2}.
\label{eq:sf}
\ee
The scintillation time  varies with frequency as $\Dtiss \propto \nu^{2/(\beta-2)}$, so the quantity $\nu^2 [\Dtiss(\nu)]^{-(\beta-2)}$, and therefore $D_{\DM}(\tau)$, is independent of frequency.

\subsection{DM and Timing Errors}

The RMS estimation error in DM,  $\sigma_{\delta\DMhat}(\tau)$, follows from 
Equation \eqref{eq:deltaDM}
\ba
\sigma_{\delta\DMhat}(\tau) = \left|\frac{r^2}{r^2-1}\right| \SFDM^{1/2}(\tau) \propto \tau^{(\beta-2)/2}
\label{eq:sigDMhat}
\ea
and the RMS error in the infinite-frequency TOA is  
\ba
\sigma_{\delta\hat t_{\infty}}(\tau) &=& K \nu_1^{-2}\,  \left|\frac{r^2}{r^2-1}\right| \SFDM^{1/2}(\tau)
		 \propto \tau^{(\beta-2)/2}.
\label{eq:sigthat}
\ea
While we consider the case where $r > 1$, the formalism presented thus far also holds for $r < 1$. Again we note that $\sigma_{\delta\DMhat}$ can be reduced by switching $\nu_1$ and $\nu_2$, as the lower frequency TOA is more sensitive to changes in DM. However, the quantity of interest, $\sigma_{\delta\hat{t}_\infty}$, will remain the same.

For $\tau$ measured in days and the scintillation time $\Dtiss$ evaluated at 1~GHz referenced to 1000~s, the DM structure function for a Kolmogorov medium is
\ba
D_{\DM}(\tau) & = & \left(1.57 \times 10^{-6}~\mathrm{pc~cm^{-3}}\right)^2 
	\nonumber \\
	&&\quad\times \left(\frac{\tau_d}{\Dtiss(1~{\rm GHz})/10^3~\mathrm{s}}\right)^{5/3}
\label{eq:sfdm_with_units}
\ea
and so the RMS error in the DM estimate is
\ba
\sigma_{\delta\widehat\DM}(\tau) &&= 1.57 \times 10^{-6}~\mathrm{pc~cm^{-3}} \left|\frac{r^2}{r^2-1}\right| 
\nonumber \\
&&\quad\times \left(\frac{\tau_d}{\Dtiss(1~{\rm GHz})/10^3~\mathrm{s}}\right)^{5/6}.
\label{eq:rms_dm}
\ea
The RMS error in the infinite-frequency TOA is  
\ba
\sigma_{\delta\hat t_{\infty}}(\tau) &\approx& \frac{6.5~{\rm ns}} {\nu_1^{2}}  \left|\frac{r^2}{r^2-1}\right| 
\nonumber \\
&&\quad\times \left(\frac{\tau_d}{\Dtiss(1~{\rm GHz})/10^3~\mathrm{s}}\right)^{5/6}.
\label{eq:rms_tinfty}
\ea

\begin{figure}[t!]
\begin{center}
\includegraphics[scale=0.42]{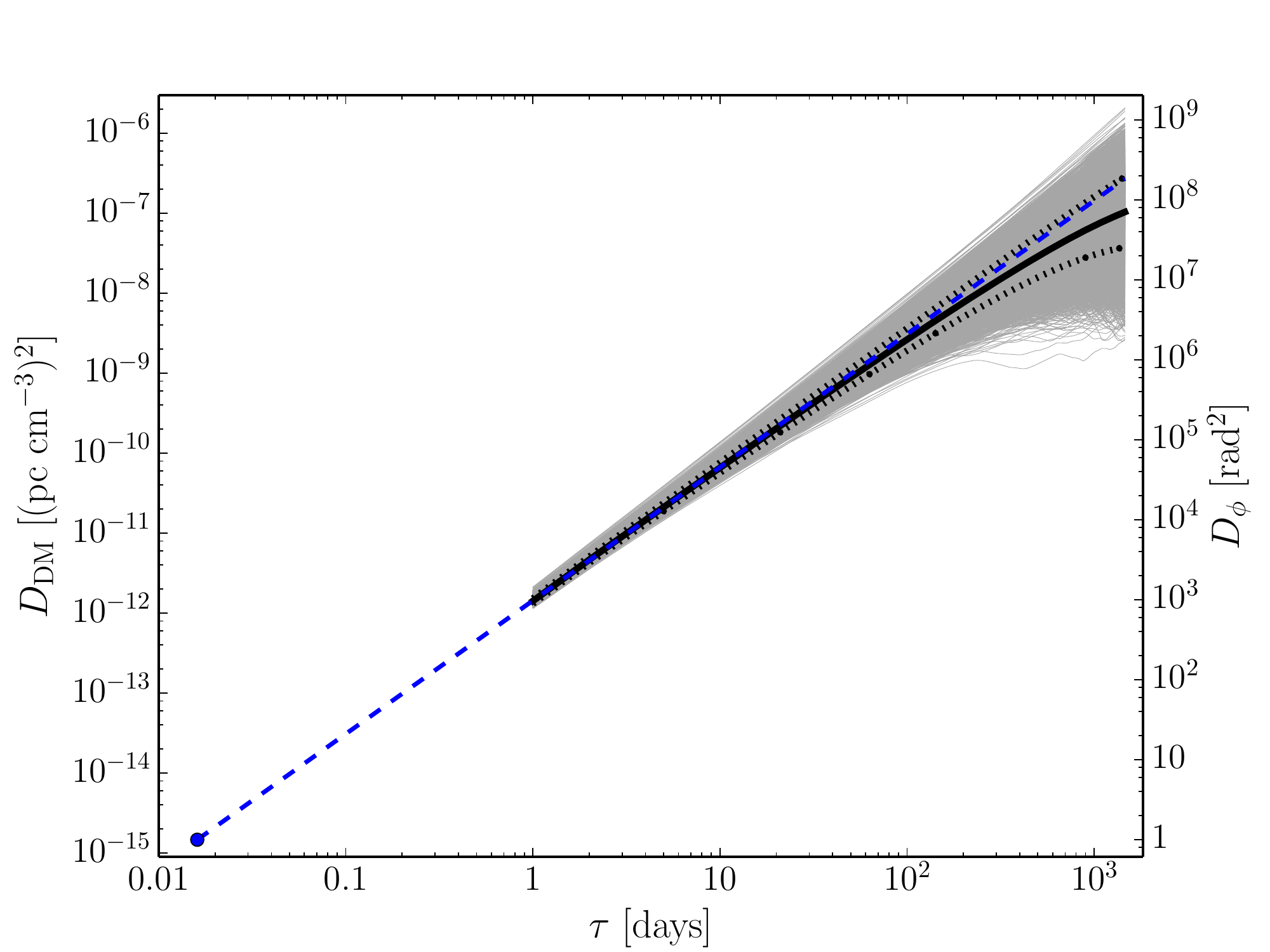} \hspace{0.5in}
\caption{\footnotesize 
Structure functions for $\beta = 11/3$. The solid, thick black line denotes the structure function (SF) geometrically averaged over $10^4$ realizations and  the dotted lines represent $\pm 1\sigma$ deviations from the mean.  The SFs for each realization are shown as thin gray lines. The filled circle at the bottom left and the dashed line show the SF extrapolated from the scintillation timescale at 1 GHz,  $\DtissGHz = 1388$ s \citep{Keith+2013}, to larger lags using Equation \eqref{eq:sfdm_with_units}. Note the small bias between the average SF and the extrapolation at large $\tau$.
\label{fig:sfs}
}
\end{center}
\end{figure}

In practice, the statistical quantities are estimated through averages over a data set of length $T$, which is typically several to many years. Time averaging does not appear in the analytical results because the statistical quantities have stationary statistics and we have assumed implicitly that the scintillation time $\Dtiss$ is epoch-dependent. However, scintillation parameters are known to vary on some LOSs (\citealt{jnk1998}, L. Levin et al., in preparation), so a more detailed treatment would average the structure function over time.

\begin{figure}[t!]
\begin{center}
\includegraphics[scale=0.42]{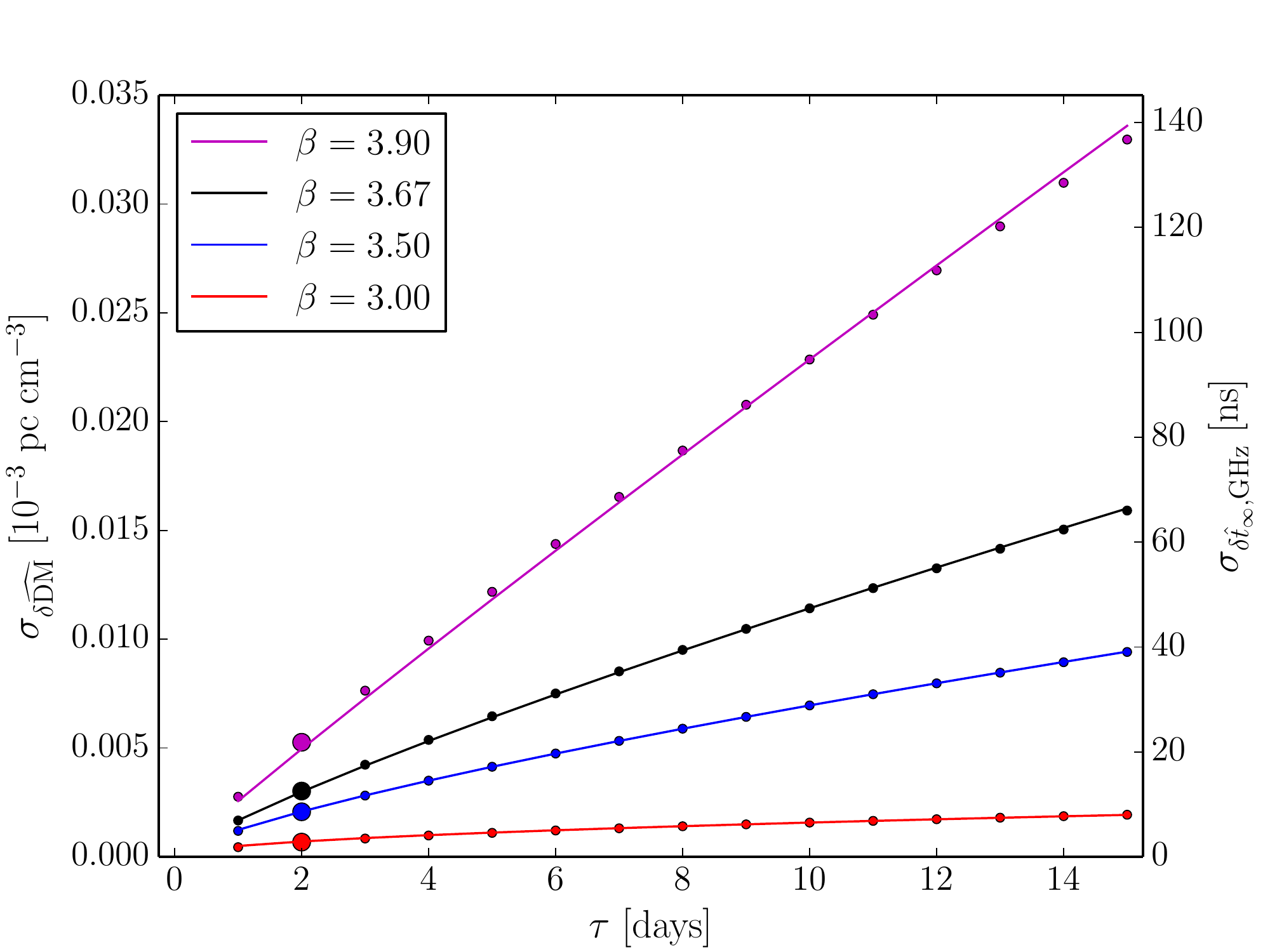} \hspace{0.5in}
\caption{\footnotesize 
$\sigma_{\delta\DMhat}$ versus multi-frequency observation offset $\tau$ for four values of $\beta$ (in the same order as the legend from top to bottom). Error bars are smaller than the plotted symbols. On the right axis we show $\sigma_{\delta\hat{t}_\infty}$ scaled to 1 GHz. The larger dots emphasize the value of $\sigma_{\delta\DMhat}$ for the median $\tau$ from the distribution shown in Figure \ref{fig:tau_histogram}. Solid lines indicate the analytic function of $\sigma_{\delta\DMhat}$ as per Equation \eqref{eq:rms_dm}. The deviations of the simulation from the analytic form come from biases in the structure function that scale with increasing $\beta$. 
\label{fig:sigma_vs_tau}
}
\end{center}
\end{figure}

\subsection{Spectral Properties}

As presented, our results do not depend on the length of the overall data span $T$ of a timing data set. If, hypothetically,  the electron density  were sampled directly to form a time series, a steep Kolmogorov-like spectrum would yield a variance that depends strongly on $T$.  Also, a Fourier-transform based power-spectral estimate would be heavily biased by spectral leakage.    

The lack of dependence of our results on $T$ follows because the observable quantity, DM, is a one-dimensional integral of the electron density and has a temporal power spectrum that is shallower than that of the electron density variations (from Equation \eqref{eq:wavenumber_spectrum}), i.e. $S_{\DM}(f)  \propto f^{-\gamma}$ with spectral index $\gamma = \beta - 1$ in the scintillation regime, as noted above.   The  DM difference $\Delta\DM(t, \tau)$ that we analyze   (e.g. Equation \eqref{eq:dDM}) is similar to a first derivative for small $\tau$. Since the Fourier transform of a first derivative in the time domain multiplies the transformed function in the frequency domain by one power of $f$, the power spectrum $S_{\DM}$ is multiplied by $f^2$ and therefore has a spectral index $\beta-3$ that is less than unity for the regime of interest.  For such shallow spectra, the variance should be independent of $T$ and spectral leakage is negligible.   We demonstrate these effects for the anticipated spectral cutoffs using simulations in the next section.   

The ensemble-average structure function can be written in terms of the power spectrum for DM, 
\be
D_{\DM}(\tau) = 4 \int df\, S_{\DM}(f) \sin^2(\pi f \tau).  
\label{eq:D_DM_to_PS}
\ee
If the integrand is dominated by frequencies where $f\tau \ll 1$ for $\tau \sim$~days, then $\sin^2(\pi f \tau) \propto (f \tau)^2$ and the structure function has a square-law form in $\tau$.

\begin{figure}[t!]
\begin{center}
\includegraphics[scale=0.42]{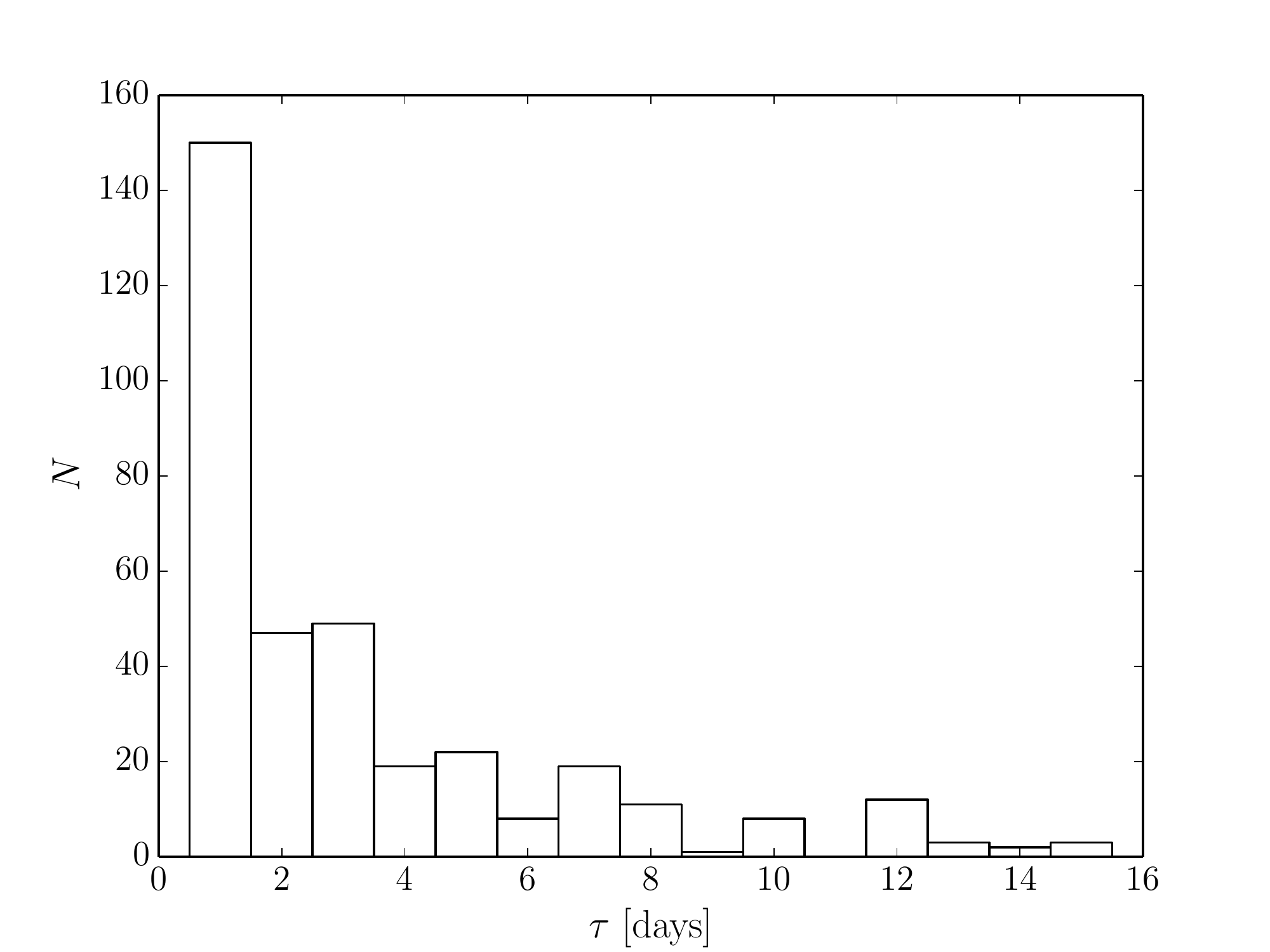} \hspace{0.5in}
\caption{\footnotesize 
Histogram of $\tau$ for the 10 pulsars observed with GBT in \cite{dfg+2013}. Here we have taken $\tau$ to be the absolute value of the time difference between observations at the two different frequency bands. The median $\tau=2$ days.
\label{fig:tau_histogram}
}
\end{center}
\end{figure}

\section{Simulations}

{We simulated DM variations consistent with the wavenumber spectrum of Equation \eqref{eq:wavenumber_spectrum} by scaling complex white noise in the frequency domain and transforming to the time domain. Wavenumber cutoffs are outside the range of corresponding timescales we probe and therefore are not implemented. We used $10^4$ realizations of  $\DM(t)$ as a red-noise process over a range of power-law indices $\gamma$ in the scintillation regime, including the Kolmogorov value $\gamma = 8/3$, each 10 years long with one-day time resolution. For specificity, we use the scintillation time scale of the MSP J1909--3744, one of the best timed objects, to set the coefficient of the phase and DM structure functions. J1909--3744 has a scintillation timescale typical of the low-DM  MSPs used for GW detection \citep{dfg+2013, Keith+2013}.  Scaled to 1~GHz, it has $\Dtiss = 1388$~s. We emphasize that no white noise has been added to model measurement errors, so we effectively assume  that DM can be recovered with no error when $\tau = 0$. We compute time series $\delta \DMhat(t,\tau)$ using Equation \eqref{eq:deltaDM} and a frequency ratio $r = 1.5~\mathrm{GHz}~/~0.8~\mathrm{GHz} = 1.875$ to match the center frequencies of observing bands at the GBT. 

Figure~\ref{fig:samplerealization} shows representative results for a single realization of $\DM(t)$ due to density variations in a medium with a Kolmogorov spectrum along the LOS coupled with changes in the LOS from relative motion. The left column shows time series $\DM(t)$ at top after the mean value has been removed and $\delta \DMhat(t,\tau=1~\mathrm{day})$ at bottom while the right column shows the respective power spectra. Note the relative flatness of the $\delta\DM$ spectrum with low spectral index $\gamma = 2/3$.

\begin{center}
\begin{deluxetable*}{lc|ccc|cc}
\tablecolumns{7}
\tablecaption{Predicted Timing Errors for Selected Millisecond Pulsars\\$\sigma_{\delta\hat{t}_\infty}$ at 1.5~GHz for $\beta=11/3,r=2$}
\tablehead{
\colhead{Pulsar} & \colhead{$\DtissGHz^{\textrm{a}}$~[s]} & \multicolumn{3}{c}{$\sigma_{\delta\hat{t}_\infty}(\tau~\mathrm{days})$~[ns]}& \colhead{$\sigma_{\mathrm{TOA}}$~[ns]} & \colhead{Observatory}\\
\colhead{} & \colhead{} & \colhead{$\tau = 1$}  & \colhead{$\tau = 3$}  & \colhead{$\tau = 5$} & \colhead{} & \colhead{}
}
\startdata
J0437--4715 & 1528 & 3 & 7 & 9 & 38$^b$ & Parkes\\
J1713+0747 & 1755 & 2 & 6 & 8 & 50$^c$ & Arecibo/GBT\\
B1855+09 & 900 & 4 & 11 & 13 & 250$^c$ & Arecibo\\
J1909--3744 & 1388 & 3 & 7 & 9 & 150$^c$ & GBT\\
B1937+21 & 201 & 15 & 37 & 47 & 35$^b$ & Parkes
\enddata
\footnotetext{All values from \cite{Keith+2013}.}
\footnotetext{Median TOA uncertainty at 1.5~GHz for 256~MHz bandwidth from \cite{Hobbs2013}.}
\footnotetext{Median TOA uncertainty at 1.5~GHz for 4~MHz bandwidth from \cite{dfg+2013}.}
\end{deluxetable*}

\end{center}

\section{Results}

Structure functions of the closely related quantities, DM and $\phi$, derived from simulations are shown  in Figure \ref{fig:sfs}  for multiple realizations with $\beta = 11/3$. The left-hand axis gives DM units and the right-hand axis phase units for $\Dtiss$ measured at 1~GHz.  The extrapolation of the structure functions from the scintillation time matches simulated results but there is a small bias between the average structure function and the extrapolation at large lags that are comparable to the length of the time series; this is a common feature of structure function estimates and underscores that structure functions need to be interpreted with caution.

\begin{figure}[t!]
\begin{center}
\includegraphics[scale=0.42]{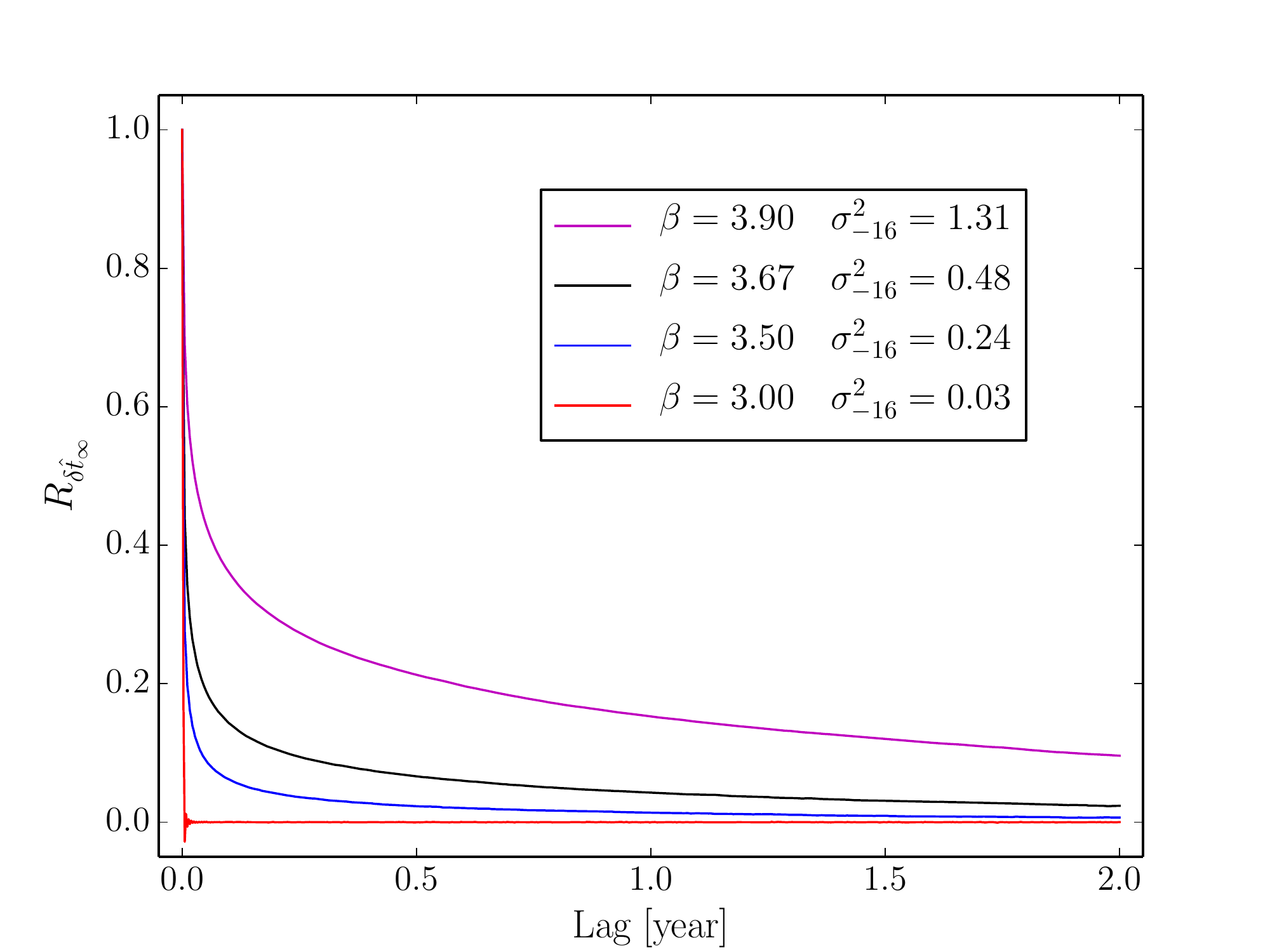} \hspace{0.5in}
\caption{\footnotesize 
Normalized, average autocorrelation functions of $\delta\hat{t}_\infty$ at 1 GHz for four values of $\beta$ (in the same order as the legend from top to bottom) and $\tau=1$ day based on $10^4$ realizations. The normalization factor is the zero lag value $\sigma^2_{\delta\hat{t}_\infty} = 10^{-16}~\sigma^2_{-16}~\mathrm{s}^2$, which is the average total variance in each time series, determined by the scintillation timescale of J1909--3744 and using Equation \eqref{eq:sf}. The particular case of $\beta=3$ shows small Gibbs ringing near the zero lag.
}
\label{fig:acfdeltaDM}
\end{center}
\end{figure}

\begin{figure}[t!]
\begin{center}
\includegraphics[scale=0.42]{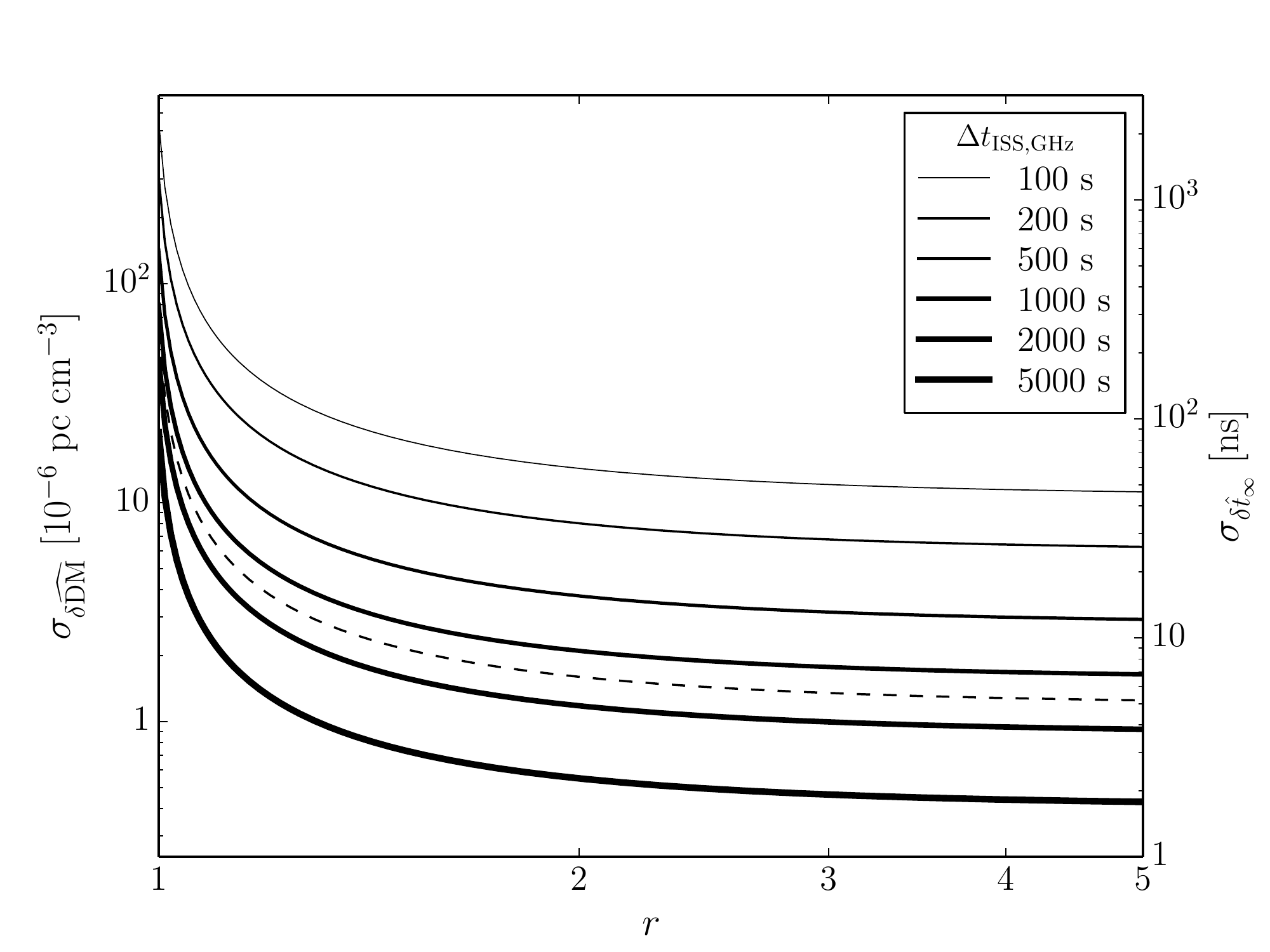} \hspace{0.5in}
\caption{\footnotesize 
Analytic evaluation of $\sigma_{\delta\DMhat}$ in Equation \eqref{eq:sigDMhat} as a function of observational frequency ratio $r = \nu_1 / \nu_2$ for different pulsar scintillation timescales $\Dtiss$, assuming a Kolmogorov medium with $\beta =11/3$ and $\tau = 1$ day. The dashed line shows the value expected for J1909--3744 \citep{Keith+2013}.
\label{fig:contour}
}
\end{center}
\end{figure}

We estimate the RMS error in DM for an observation gap $\tau$, $\sigma_{\delta\DMhat}(\tau)$, by averaging over the $10^4$ realizations. Figure \ref{fig:sigma_vs_tau} shows $\sigma_{\delta\DMhat}$ vs $\tau$ for several nominal values of $\beta$ near the Kolmogorov case. Even though the analytic value in Equation \eqref{eq:rms_dm} matches our simulations well, slight biases will cause mismatches that scale with increasing $\beta$  \citep{Rutman1978,cd1985}. We expect the simple $\tau^{\beta-2}$ scaling only in the scintillation regime for $2 < \beta < 4$ but we also need $\tau$ to be at  an intermediate value such that $f_2^{-1} \ll \tau \ll f_1^{-1}$ where $f_1$ is the lower frequency cutoff and $f_2$ is the upper frequency cutoff. The simple $\tau^{\beta-2}$ scaling is perturbed once lags become comparable to the data span length. The perturbation is why we see a slight deviation of the structure function from the simple scaling law for larger lags. Once $\beta > 4$, the structure function will have a  square-law form $\propto \tau^2$ for most lag values.  Note that for $\beta < 2$ the simple scaling no longer holds. 

Figure \ref{fig:tau_histogram} shows a histogram of offsets between 1.5 GHz and 0.8 GHz observations from GBT observations of 10 pulsars in the NANOGrav five-year data set \citep{dfg+2013}. The median $\tau = 2$~days implies $\sim 13$ ns of timing error for a pulsar such as J1909--3744 at 1 GHz, a non-negligible fraction of the five-year weighted timing RMS of 38 ns \citep{dfg+2013}.

It is instructive to diagnose the amount of temporal correlation expected in the time series for $\delta\hat{t}_\infty$. 
Autocorrelation functions (ACFs) of $\sigma_{\delta\hat{t}_\infty}$ averaged over realizations are shown in Figure \ref{fig:acfdeltaDM} for four values of $\beta$. The amplitudes of the ACFs are set by the extrapolation of the DM structure function from the scintillation timescale. The correlation timescale, estimated by the width of the ACF, increases with $\beta$, indicating that red noise is introduced into timing residuals but with a spectral index (i.e. from the power spectrum) that is no more than unity. 

Figure \ref{fig:contour} shows $\sigma_{\delta\DMhat}$ (see Equation \eqref{eq:rms_dm}) as a function of the frequency ratio $r$ for various scintillation timescales $\Dtiss$ of a given pulsar. An increase in the frequency ratio $r$ will reduce the amount of error expected from non-simultaneous observations. The DM and TOA errors asymptote to constant values for large $r$ though most of the reduction in these errors is obtained for $r = 2$. These errors can also be reduced by decreasing the number of days of separation between observations $\tau$. We re-emphasize that our analysis excludes measurement errors.

In reality, errors in DM estimation from additive noise can be improved by an increase in $r$, up to a point dependent on the pulsar's intrinsic frequency spectrum, regardless of the systematic error from non-simultaneous observations. Therefore, timing campaigns need to optimize the two kinds of error with respect to a choice of $r$ on a pulsar-by-pulsar basis. 

We show the analytic values of $\sigma_{\delta\hat{t}_\infty}$ scaled to 1.5~GHz for several of the best-timed MSPs in Table 1. Errors on the order of 10~ns can be expected at the level of our target timing precision for GW detection and a significant fraction of the TOA uncertainty due to radiometer noise even as telescope backends improve. As a comparison, we list the smallest published median template fitting errors, $\sigma_{\mathrm{TOA}}$, across bands centered close to 1.5~GHz. Values between observatories scale approximately as $B^{-1/2}$ for a bandwidth $B$, though scintillation can change these numbers somewhat. We see that for the selected best-timed MSPs, errors from non-simultaneous observations become non-negligible, even for small values of $\tau$, and we should attempt to reduce these errors accordingly. When $\tau$ cannot be zero (e.g. due to telescope constraints), one alternate method of error reduction is to add additional DM measurements by increasing the number of observing epochs.

\section{Extension to N-Point Sampling}

We can extend our analysis from two-point sampling of DM to $N$-point sampling. For simplicity, we start with the three-point case and assume the three TOA measurements are spaced by $\tau$ days of separation, i.e. $t_1 = t - \tau, t_2 = t, t_3 = t + \tau$, at frequencies $\nu_1$, $\nu_2$, and $\nu_3$, respectively. We can define the estimated DM fluctuation between epochs $t_i$ and $t_j$ to be analogous as previously defined,
\be
\DMhat_{ij}(t,\tau) = \frac{\Delta t_i - \Delta t_j}{K(\nu_i^{-2} - \nu_j^{-2})}.
\ee
If we reference the DM measurements to time $t$, we find the overall DM estimate to be
\be
\DMhat(t,\tau) = \frac{1}{2}\left[\DMhat_{21}(t,\tau) + \DMhat_{32}(t,\tau)\right],
\ee
the average of both pairs of measurements.

While the frequency ratio used for $\DMhat_{21}$ and $\DMhat_{32}$ need not be the same, we will assume $\nu_1 = \nu_3$, so that for $r_{ij} = \nu_i / \nu_j$ we have $r_{21} = r_{23} = r$. We can then write the difference between the true and estimated DM as
\be
\delta \DMhat(t,\tau) = \frac{1}{2}\left(\frac{r^2}{r^2-1}\right)\Delta^{(2)}\DM(t, \tau),
\label{eq:deltaDM3pt}
\ee
where the superscript (2) denotes the second increment of DM, defined to be
\be
\Delta^{(2)}\DM(t,\tau) \equiv \DM(t-\tau) - 2\DM(t) + \DM(t+\tau).
\label{eq:d2DM}
\ee
The second-order structure function, $D_{\DM}^{(2)}(\tau) = \left\langle\left[\Delta^{(2)}\DM(t,\tau)\right]^2 \right\rangle$, allows us to write the RMS estimation error,
\be
\sigma_{\delta\DMhat}^{(2)}(\tau) = \frac{1}{2}\left|\frac{r^2}{r^2-1}\right| \left[D_{\DM}^{(2)}(\tau)\right]^{1/2}
\label{eq:sig2DMhat}
\ee

Similarly to the two-point case, the ensemble-average, second-order structure function can also be written in terms of the power spectrum for DM, 
\be
D_{\DM}^{(2)}(\tau) = 16 \int df\, S_{\DM}(f) \sin^4(\pi f \tau).  
\label{eq:D_DM_to_PS_3pt}
\ee
Combining Eqs.~\eqref{eq:sigDMhat} and \eqref{eq:sig2DMhat} with \eqref{eq:D_DM_to_PS} and \eqref{eq:D_DM_to_PS_3pt}, and converting to time units, we can solve for the ratio of RMS estimation errors between the three-point and two-point sampling cases,
\ba
R(\tau)= \frac{\sigma_{\delta\hat{t}_\infty}^{(2)}\!(\tau)}{\sigma_{\delta\hat{t}_\infty}^{(1)}\!(\tau)}&=&\frac{1}{2}\!\left[\frac{D_{\DM}^{(2)}(\tau)}{D_{\DM}^{(1)}(\tau)}\right]^{1/2}\nonumber\\
&=&\left[\frac{\int df\, S_{\DM}(f) \sin^4(\pi f \tau)}{\int df\, S_{\DM}(f) \sin^2(\pi f \tau)}\right]^{1/2}. 
\label{eq:RMS_ratio} 
\ea
For $S_{\DM}(f) \propto f^{-\gamma}$ in the scintillation regime $1 < \gamma < 3$, we can solve for this ratio (now labeled with subscript $\gamma$) exactly as $R_{\gamma}(\tau) = \sqrt{1-2^{\gamma-3}}$, for values of $\tau$ that satisfy the inequalities in \S 5 and as the lower frequency cutoff $f_1 = 1/T$ tends to zero. The Kolmogorov case implies $R_{8/3} \approx 0.45$ and three-point sampling allows for a greater than factor of two improvement over two-point sampling on the RMS estimation error. Since the second-order structure function removes linear trends from the time series, whereas the first-order structure function only removes constant terms, $R$ will decrease even further if $\DM(t)$ is slope-dominated as is the case for even steeper wavenumber spectra. 

For an arbitrary number of epochs $N$ used in DM estimation, $N$-point sampling will involve an ($N$-1)-order structure function. The increased cost of observing time will yield diminishing improvements in the RMS error. Following Equation \eqref{eq:RMS_ratio} for the reduction in RMS error, the integrand in the numerator will contain increased powers of $\sin^{2(N-1)}(\pi f \tau)$. Since the ($N$-1)-order increment will have a power spectrum with associated power-law index $\gamma-2(N-1)$ for small values of $f \tau \ll 1$, increasing the number of sampling epochs $N$ will cause the spectral index to become more positive, i.e. the spectral slope will become more positive, and the ratio of RMS estimation errors between the $N$-point and the two-point cases will grow. For a Kolmogorov spectrum, we find that the three-point case results in the greatest reduction in error. For steeper wavenumber spectra, increased sampling may become important.

At a telescope like the GBT, the cost of implementing a three-point sampling scheme involves $\sim50\%$ added time to observing at one frequency band over each set of observations, but could yield an important reduction in the timing noise budget for MSPs like J1713+0747 and J1909--3744. A high-low-high frequency observing scheme at the GBT will mean both a doubling of higher signal-to-noise TOAs measured at 1.5~GHz as well as an improvement in DM estimation as $r < 1$. As in the two-point case, $\sigma_{\delta\DMhat}$ can be reduced for $r < 1$ but $\sigma_{\hat{t}_\infty}$ cannot.  While we only consider an extension to three-point sampling here, arbitrarily increasing the number of sampling days should improve the DM estimate, though with a further increased cost of observing time. The best-case scenario involves the construction of new high-sensitivity, ultra-wideband receiver systems spanning enough bandwidth to allow for accurate enough DM estimation that would eliminate the need for multi-epoch observations altogether.

\section{Discussion And Conclusions}

Our simulations of non-simultaneous, multi-frequency observations indicate a lower bound to timing errors on the order of $\sim10$s of nanoseconds for LOSs with scintillation timescales comparable to those of current MSPs sampled in PTAs.   The timing error results from mis-estimation of the DM and any additional measurement errors will increase the DM errors and therefore the TOA uncertainty further. For the Kolmogorov case, we find a  ``pink'' noise spectrum for the time series of DM errors proportional to $f^{-2/3}$. Any red noise present in timing residuals can affect the sensitivity of a PTA to GWs. For identical observational parameters in a timing program, these campaigns will be limited by the induced DM measurement error largely related to the scintillation timescale for a given pulsar. TOA errors scale as $\sigma_{\delta\DMhat} \propto \Dtiss^{-5/6}$, so timing measurements of pulsars with larger $\Dtiss$ will have smaller errors.

While we analyze DM variations with power-law wavenumber spectra and wavenumber cutoffs outside of the corresponding timescales we probe, observed interstellar DM variations have a minimum characteristic timescale that is determined by spatial smoothing from scattering and is equal to the refraction timescale \citep{cs2010},  $t_r \sim 2.4~\mathrm{days}~(D/D_s)(\nu_{\mathrm{GHz}} / \Delta\nu_{\mathrm{ISS},0.01}) (\Dtiss / 1000~\mathrm{s}),$  where $D$ is the distance to the pulsar, $D_s$ is the distance between the pulsar and the scattering screen, $\Delta\nu_{\mathrm{ISS},0.01}$ is the scintillation bandwidth is in units of 10 MHz. For J1909-3744, the scintillation bandwidth is 37~MHz at 1.5~GHz \citep{Keith+2013}, which means that the refraction timescale is on the order of a day unless $D_s \ll D$, changing the effective higher wavenumber spectrum cutoff. Because $t_r \propto \nu / \Delta\nu_{\mathrm{ISS}} \propto \nu^{-17/5}$,  the relevant smoothing time is from the {\it higher} of a pair of frequencies. The difference in smoothing at the two frequencies gives rise to other effects that are discussed in a separate paper (J. M. Cordes et al. in preparation). 

As we discover MSPs farther out in the Galactic plane, we expect an increase in DM along these LOSs. While MSPs with higher DMs have the potential to be suitable for inclusion into a PTA, mitigation of increased scattering effects may prove challenging. We expect that $\Dtiss \propto \DM^{-3/5}$ for a homogenous, Kolmogorov medium \citep[e.g. Equation (46) of][]{cl1991}. With decreasing $\Dtiss$, more stringent constraints on $\tau$ will become important for the population of newer, more distant pulsars.

Equation \eqref{eq:toa_error} describes the timing error associated with any DM differences between observations. While we have only considered electron-density variations from fluctuations in the ISM so far, local variations, such as in the ionosphere, can also add to the TOA error. Ionospheric changes in electron density correlate with incident flux from the Sun. Daily changes due to Earth's rotation, yearly changes due to Earth's orbit, and eleven-year cycles due to solar magnetic activity, are all observed, leading to electron-density variations that can vary by up to $\sim 10^{-5}~\mathrm{pc\;cm^{-3}}$, or a TOA uncertainty of $\sim 40~\mathrm{ns}$ at 1~GHz, on any of these timescales \citep{hr2006}. Extreme solar events can produce a bigger effect. The observed electron density will also increase for LOSs that pass through large zenith angles. The uncertainty in DM can increase even for small values of $\tau$ if the pulsar is observed at widely different hour angles. Since the ionosphere varies spatially across the globe as well as temporally, DM differences, and therefore associated TOA errors, can potentially exist in simultaneous, multi-frequency observations for separate telescopes at a level larger than the RMS error in Equation \eqref{eq:rms_dm}. We suggest that the best practice is for observations to be spatially coincident as well as simultaneous.

Equation \eqref{eq:rms_dm} can be used to quantify the tolerance level in non-simultaneous observations for minimal acceptable levels of noise. Observations requiring DM correction should ensure a frequency ratio $r \gtrsim 2$, which is well within the goal of future wideband timing systems. Timing campaigns should strive for same day observations, though alternatively, measurements with multiple ($> 2$) observations per epoch may partially improve DM estimation and the overall TOA uncertainty. Errors from non-simultaneous multi-frequency measurements may become a large contribution to the total timing noise budget for pulsars with the highest-timing precision, for pulsars observed with next-generation telescopes with improved sensitivity, and for PTAs as a whole since sensitivity is often dominated by the best-timed pulsars in the array.

\acknowledgments

Work on pulsar timing at Cornell University is supported in part by NSF PIRE program award number 0968296.

\end{document}